\documentclass[sigconf]{acmart}
\usepackage[svgnames]{xcolor}

\usepackage{listings}
\usepackage{xcolor}

\definecolor{codebg}{RGB}{248,248,248}
\definecolor{codegreen}{RGB}{0,128,0}
\definecolor{codepurple}{RGB}{170,0,255}
\definecolor{codeblue}{RGB}{0,0,255}

\lstdefinestyle{fancypython}{
    language=Python,
    backgroundcolor=\color{codebg},
    commentstyle=\color{codegreen}\itshape,
    keywordstyle=\color{codeblue}\bfseries,
    stringstyle=\color{codepurple},
    numberstyle=\tiny\color{gray},
    basicstyle=\ttfamily\footnotesize,
    breaklines=true,
    numbers=left,
    numbersep=8pt,
    showstringspaces=false,
    tabsize=4,
    frame=single,
    morekeywords={self,True,False,None},
    morestring=[b]',
    morestring=[b]",
    morecomment=[l]{\#}
}

\usepackage{balance}
\AtBeginDocument{%
  }

\copyrightyear{2026}
\acmYear{2026}
\setcopyright{cc}
\setcctype{by}
\acmConference[SIGCSE TS 2026]{Proceedings of the 57th ACM Technical Symposium on Computer Science Education V.2}{February 18--21, 2026}{St. Louis, MO, USA}
\acmBooktitle{Proceedings of the 57th ACM Technical Symposium on Computer Science Education V.2 (SIGCSE TS 2026), February 18--21, 2026, St. Louis, MO, USA}
\acmPrice{}
\acmDOI{10.1145/3770761.3777313}
\acmISBN{979-8-4007-2255-4/2026/02}

\begin{document}

\title{LEAP – Live Experiments for Active Pedagogy}




\author{Sumedh Karajagi}
\authornote{Both authors contributed equally to this research.}
\orcid{0009-0001-6099-6634}
\affiliation{%
\institution{California Academy of Mathematics and Science}
  \city{Carson, California}
  \country{USA}}
\email{sumedh1karajagi@gmail.com}

\author{Sampad Bhusan Mohanty}
\authornotemark[1]
\orcid{0000-0002-1161-9656}
\affiliation{%
  \institution{University of Southern California}
  \city{Los Angeles, California}
  \country{USA}}
\email{sbmohant@usc.edu}

\author{Bhaskar Krishnamachari}
\orcid{0000-0002-9994-9931}
\affiliation{
\institution{University of Southern California}
  \city{Los Angeles, California}
  \country{USA}}
\email{bkrishna@usc.edu}

\renewcommand{\shortauthors}{Sumedh Karajagi, Sampad Bhusan Mohanty, and Bhaskar Krishnamachari}

\begin{abstract}
  Interactive computational environments can help students explore algorithmic concepts through collaborative hands-on experimentation. However, static and instructor controlled demos in lectures limit engagement. Even when interactive visualizations are used, interactions are solely controlled by the instructor, leaving students as passive observers. In addition, the tools used for demonstration often vary significantly, as they are typically developed by individual instructors. Consequently, the visualizations remain confined to a single classroom, rather than being shared and adapted across courses or reused by other instructors. To address this gap and foster active engagement in live classrooms, we present a lightweight and seamless software framework named LEAP for developing interactive computational lab exercises using a simple idea: remotely callable instructor-defined functions. Using API endpoints and a provided client, students can discover and then call instructor defined functions remotely from their coding environment using scripts or interactive notebooks. Each function call is time-stamped and persistently logged in a database, allowing real-time visualization of participation, diverse solution paths, common pitfalls, and live feedback through collaboration, gamification, and quizzes. Labs are packaged as self-contained folders, each containing their own remotely callable functions. We provide example labs to demonstrate applications relevant for numerical analysis, machine learning, algorithms courses and mention some in electrical engineering (EE), economics, and physics. These capabilities enhance engagement and provide instructors with actionable insights into learning processes. With a standardized lab format and an online directory for community-contributed labs, we aim to foster a global ecosystem for exchanging and expanding interactive pedagogy enabled by LEAP.
\end{abstract}





\maketitle

\section{Introduction}
Active learning consistently improves outcomes in education, with flipped classrooms, problem-based learning, gamification, and collaboration enhancing both engagement and performance \cite{mti8060050}. Collaborative approaches, including classroom-wide activities, also improve course success compared to independent work \cite{10.1145/3372782.3406254}\cite{article}. Yet many computer science (CS) courses still rely on static lectures and instructor-side computing environments for classroom demonstrations where students are passive participants. At best, student interactions are usually directed toward the instructor, and collaborative interactions between students are limited during demonstrations. Although software like \textit{dpvis} improved student comprehension of recursion and tabular methods through dynamic programming visualization \cite{10.1145/3641554.3701865} and while IPython notebooks demonstrated how interactive coding and visualization can support inquiry-based learning in numerical methods \cite{ketcheson-proc-scipy-2014}, they do not facilitate collaborative student interactions for in class demonstrations.

Motivated by this, we present an open-source framework called LEAP. LEAP demonstrations (or labs) are folders with all necessary resources needed for deployment on a local or cloud server, containing one or more experiments and a collection of remotely callable functions. Students interact with the system through a web browser or via the LEAP client in their preferred programming language. LEAP emphasizes simplicity: the client provides only two RPC API endpoints for students — i) remote function discovery and ii) remote function calling. The function discovery service allows available remote functions, their signatures, and documentation to be discovered. The other endpoint is used by students to call remote functions with suitable arguments (via web UI or LEAP client). For every remote function call, the name of the function invoked, the arguments, and the return value are stored along with timestamp and the student ID of the student who called it in a database. These logs can be queried in real-time for many important and creative tasks, some of which could be visualizing class participation, experiment progress, and a fun leader board for friendly competition. The simple RPC paradigm allows for collaborative classroom activities and gamification of the experiments in the sense that the state of the demonstration and its progress is modifiable by any participating student and the live visualizations can steer how students engage with the experiment. This would significantly enhance student engagement, experience, and provide instructors with actionable real-time insights. Beyond CS classrooms, LEAP enables designing live quizzes and questionnaires in HTML/Markdown and real-time statistics on the answers collected. In situations where computational devices are not available for every student, as is true in many poor and developing countries, low cost user input devices can be used instead. We are working on a prototype open-source hardware user input device for use with LEAP. 

\section{System Architecture and Implementation}
The framework consists of four important parts.  \\
\textbf{1) Labs:}
Content is organized into two subdirectories - \texttt{ui/} contains files required for interactive web interfaces, and \texttt{funcs/} contains Python functions to be remotely callable. Labs may include post-experiment quizzes, placed at \texttt{ui/quizzes}. \\
\textbf{2) Core Services:} LEAP server exposes API endpoints providing four core services:
\textbf{/discover} for function introspection,
\textbf{/call} for invoking exposed remote functions,
\textbf{/logs} for querying logs,
\textbf{/labs, /admin} for managing sessions, experiments, UI updates, authentication, administrative tasks.
Security and privacy through PBKDF2 credential hashing and session-based access control with timeouts. Strict data isolation preserves lab-level privacy boundaries.
\\
\textbf{3) Client:}
Students either use the web UI or the client library to discover and interact with the experiment via the functions exposed by the instructor for various tasks. LEAP Client is available in Python and will be made available for other languages soon. \\
\textbf{4) Analytics:} Provides analytics for actionable insights on student engagement, performance, pool of problem-solving strategies,  completion times, leader-board, compute load, network load, etc.\\

\section{Example Applications}
LEAP supports active and collaborative learning across computational courses like numerical analysis (NM), optimization (OPT), graph algorithms, and machine learning (ML). Broader courses can also benefit from the live quizzes and real-time feedback enabled by LEAP. Planned extensions of LEAP include integration with hardware-based labs in EE and physics—for example, using LEAP to control propellers using PID algorithms. Another planned feature to let students share and invoke each other's remote functions. \\
\textbf{ML/NMA/OPT:} 
In a gradient-descent lab, the instructor exposes the gradient of a function \texttt{\_f(x)} as \texttt{gradient(x,y)}.
\begin{lstlisting}{python}
import jax; import jax.numpy as jnp
def _f(x, y):  # not exposed for RPC due to leading "_"
  return (((x-20)**2 + 10*20**2) * (5*(x+20)**2 + (y+20)**2) )/100
def gradient(x, y): # exposed
    x_jax,y_jax= float(x), float(y)
    g = jax.grad(_f, argnums=(0, 1))(x_jax, y_jax)
    return (float(g[0]), float(g[1]))
\end{lstlisting}
Students are tasked to implement a version of gradient descent (GD) and try their implantation on the exposed function with random initial points; the trajectory of the points traversed by their implementation is logged and available for visualizations and optional analysis against an instructor verified implementation.

\begin{figure}[h]
    \centering
    \includegraphics[width=0.9\linewidth]{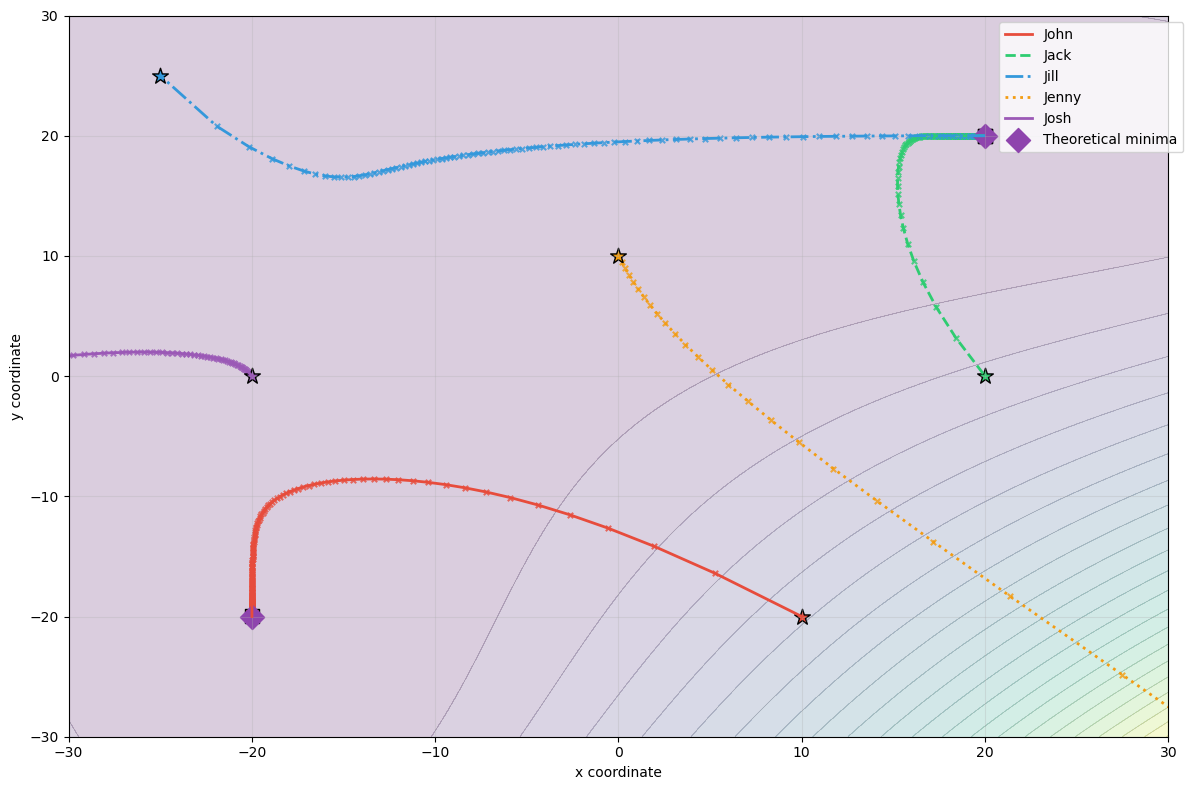}
    \vspace{-1em}
    \caption{Gradient Descent Trajectories}
    \label{fig:gd}
\end{figure}

\begin{lstlisting}{python}
# student side implementation of grad descent
import leap.client as RPCClient; import numpy as np
client = RPCClient.init(server ='http://serverIP',student_id='s001')
def gdescent(df, x, lr=1e-3, iters=300):
    for i in range(iters):
        x = x - lr * np.array(df(*x))
    return np.array(x)
x0 = np.array([10.0, 5.0])  # Starting point
gdescent(client.gradient, x0, lr=1e-3, iters=300)
\end{lstlisting}
Figure \ref{fig:gd} is plotted from a real-time dashboard implemented using \textit{marimo} embedded inside LEAP, using demonstration data from four hypothetical students. This facilitates students to visualize various trajectories taken by classmates leading to the discovery/realization that there are multiple local minima. Another important consequence of this is that common mistakes and modes of failure made by a large number of students can easily be discovered. In the figure, Jenny and Josh made the error of adding the gradients instead of subtracting from the current estimate of the minima leading to gradient ascent. This kind of learning from their classmates mistakes is something that is lacking in current pedagogy and highlights one of the many insights enabled by LEAP.
\vspace{-1em}
\vspace{-1em}
Other example labs currently provided in LEAP include Monte-Carlo sampling method integration, where students help in crowd-sourcing random samples (via a web UI). In the Numerical Integration lab, the class can be grouped, enabling collaboration, so that each group implements a different algorithm; for example, backward and forward Euler methods. The results of the various groups can then be displayed on top of each other for comparison (friendly competition). Similarly, the bisection method, secant method, and newton's method can be assigned to three different groups in the root finding lab.  In graph traversal lab, a web UI visualizes step by step walk on the node visits based on the students implementation of a search algorithm (BFS/DFS/etc.). To be implemented is the genetic algorithm lab, that allows students to use web UI to pick two individual solutions from the solution pool, breed an offspring, mutate them and release back into the pool and a leader-board that shows the students whose solutions have the best fitness.

\vspace{-0.5em}
\bibliographystyle{ACM-Reference-Format}
\bibliography{main}


\end{document}